\documentclass[conference]{IEEEtran} 
\usepackage{amsmath}
\usepackage{cite}
\usepackage{amsfonts}
\usepackage{amssymb}
\usepackage{graphicx}
\usepackage{subfigure}
\usepackage{algorithm}
\usepackage{algpseudocode}
\usepackage{mathtools}

\usepackage{textcomp}
\usepackage[export]{adjustbox}
\usepackage{float}
\usepackage{booktabs}
\usepackage{multicol}
\usepackage{xparse}
\usepackage{color,soul}
\usepackage[bottom]{footmisc}
\usepackage[binary-units=true]{siunitx}
\DeclareSIUnit{\belmilliwatt}{Bm}
\DeclareSIUnit{\dBm}{\deci\belmilliwatt}
\DeclareSIUnit[per-mode=symbol,per-symbol=p]{\Bps}{\byte\per\second}
\sethlcolor{yellow}
\usepackage{algpseudocode}

\usepackage{hyperref}

\makeatletter
\def\BState{\State\hskip-\ALG@thistlm}
\makeatother

\begin{document}

\title{Remote Health-Monitoring of First Responders \\ over TETRA Links
}

\author{
\IEEEauthorblockN{Hossam Farag\IEEEauthorrefmark{1}, Aleksandar Vuji\' c\IEEEauthorrefmark{2}, Milo\v s Kosti\' c\IEEEauthorrefmark{3}, Goran Bijeli\' c\IEEEauthorrefmark{4}, and \v{C}edomir Stefanovi\'{c}\IEEEauthorrefmark{1}}
\IEEEauthorblockA{
\IEEEauthorrefmark{1}Department of Electronic Systems, Aalborg University, Denmark\\
\IEEEauthorrefmark{2}TeleGroup, Serbia \\
\IEEEauthorrefmark{3}Tecnalia Serbia, Serbia\\
\IEEEauthorrefmark{4}Tecnalia Research and Innovation, Spain\\
Email: hmf@es.aau.dk,aleksandar.vujic@telegroup-ltd.com,\{milos.kostic,goran.bijelic\}@tecnalia.com,cs@es.aau.dk}}

	\maketitle
	\begin{abstract}
In this paper, we present a system for remote health-monitoring of first responders over TETRA radio links. The system features a smart garment that periodically records and sends physiological parameters of first responders to a remote agent, which processes the recordings and feeds back the health-status notifications and warnings in the form of electrotactile stimuli. The choice of TETRA as the connectivity solution is driven by its routine use by first responders all over the world, thus representing a convenient and economically-effective connectivity basis. Although the support for data communications in TETRA is rather limited and in practice reduced to the Short Data Service, we show that TETRA can serve the intended purpose in the considered scenario, achieving tolerable delay and message-loss performance. Moreover, when the system is examined and optimized in terms of the peak Age-of-Information, a metric suitable to characterize the quasi-periodic nature of the monitoring process, its performance becomes rather favorable, enabling timely monitoring of the first responders' health status. 

	\end{abstract}
	
\section{Introduction}

The focus of this paper is on a technological solution that falls at the intersection of of the e-health and public safety verticals. Specifically, we assess the solution for remote monitoring of the physiological status of the first responders, e.g., firemen, mountain rescue teams, coastal guards, etc.
These emergency services have a long history of using professional mobile radio-systems, i.e., public-safety networks, the most prominent example of which is the Terrestrial Trunked Radio (TETRA)~\cite{TETRA}.
Although TETRA systems were primarily designed for voice communications, 
their penetration and routine use present a clear incentive to also exploit it as the connectivity enabler for novel use cases and applications.
The reasons are numerous and, besides the ones related to convenience, include practical ones, e.g., equipping a first responder with a single communication device limits the physical burden and the overall system complexity, as well as economical ones, i.e., lowering capital and operating costs.

On the other hand, remote-health monitoring involving automated data analysis and decision making is a showcase example of an IoT system. 
In such system, the communication patterns both in the uplink (delivery of the monitored physiological parameters of the first responders) and in the downlink (notifications and warnings about physiological strain) are sporadic and consist of short packets, such that the required data rates can be expected to be rather modest.
However, the support for data communications in TETRA systems is limited~\cite{TETRA-AIR}.
Moreover, connecting external data devices over TETRA network is envisioned over the Peripheral Equipment Interface (PEI)~\cite{TETRA-PEI}, which, in practice, typically provides access only to the TETRA Short Data Service (SDS)~\cite{TETRA-AIR}.
SDS takes place over the shared signalling channel and is fairly similar to short message service in GSM, offering of transmission of messages that are up to 2047 bits in size.
These constraints pose valid concerns when it comes to the realization of IoT applications over TETRA networks.

This paper investigates the use of TETRA system as a connectivity enabler for remote health-monitoring of first responders in field deployments. 
Our work is motivated by a research and development project~\cite{SixthSense}, from which the communication requirements were derived.
Using the standardized evaluation approach~\cite{TETRA, TETRA-AIR} and assuming a realistic number of team members, we show that TETRA SDS service offers a tolerable performance in terms of message delay and failure probability and provide insights on the impact of the system parameters on these metrics.
Moreover, as the considered application involves quasi-periodic message exchanges, we assess the performance in terms of the peak Age-of-Information (PAoI)~\cite{AoI}.
We show that, when the TETRA system is optimized according to the possibilities allowed by the standard~\cite{TETRA-AIR}, it achieves a rather favorable PAoI performance, enabling timely monitoring of the first responders' health status.

The remainder of the text is organized as follows.
Section~\ref{sec:background} presents a brief overview of the public safety networks and of related work. 
Section~\ref{sec:architecture} introduces the system architecture and the communication requirements of the remote-health monitoring application.
Section~\ref{sec:TETRA} elaborates the TETRA SDS service.
Section~\ref{sec:evaluation} presents the evaluation methodology and the results.
Finally, Section~\ref{sec:conclusions} concludes the paper.

\section{Background and Related Work}
\label{sec:background}

    \begin{figure*}[t!] 
		\centering
		\includegraphics[width=0.9\textwidth]{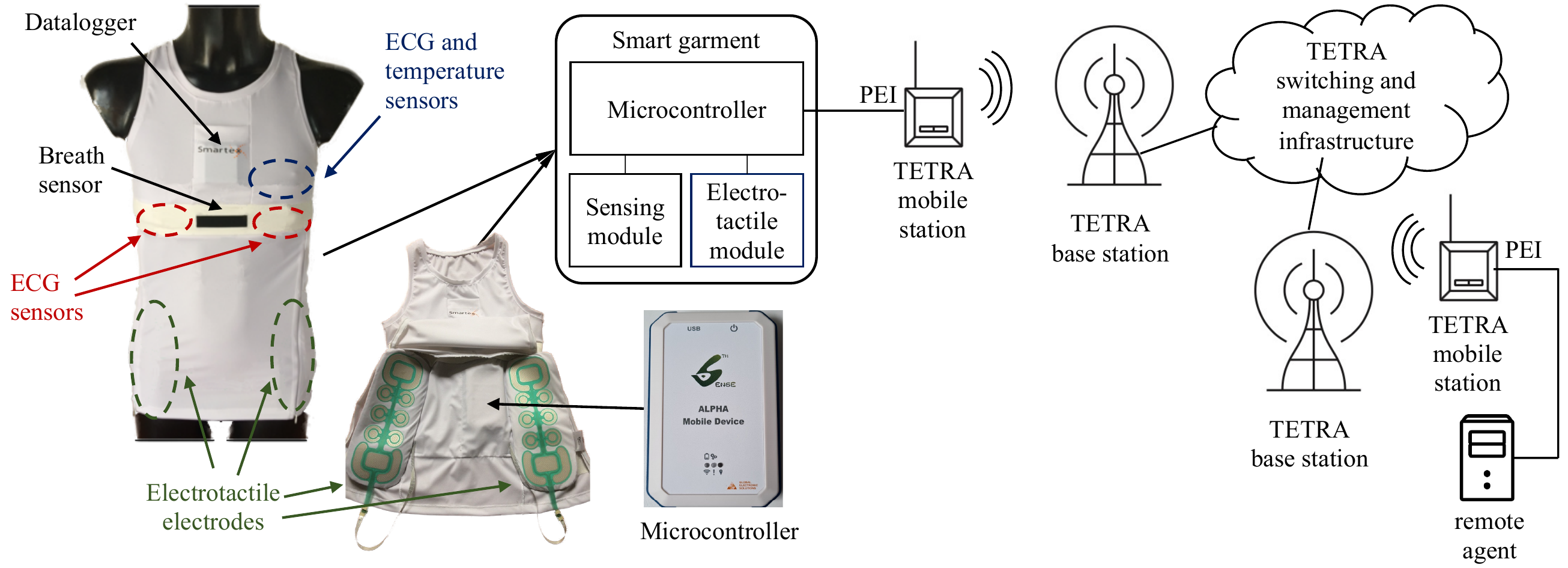}
		\caption{Block diagram of the considered remote health monitoring system.}
		\label{fig:system}
	\end{figure*}

TETRA is a European Telecommunications Standards Institute (ETSI) standard for professional mobile radio systems. 
The set of TETRA services include voice calls (group or private), circuit-switched data, packet-switched data and SDS. 
The services are characterized by mission-critical performance such as fast call setup fast message transmission, priority-based call handling, advanced encryption, and authentication.
It is widely used by government agencies, emergency services, military and for railway communications. 
It supports both the trunk mode operation (TMO), in which the service takes place over the switching infrastructure composed of the base stations, as the direct mode operation (DMO), in which the TETRA terminals directly communicate to each other.
TETRA operates (primarily) in the 400~MHz band, reaching long ranges at the cost of low data rates -- the maximum achievable data rate in the original release of the standard is 28.8~kbit/s~\cite{TETRA-AIR}.
The later version of the standard introduced TETRA Enhanced Data Services (TEDS)~\cite{TEDS}, offering data rates of up to 500~kbit/s.
The original standard defines PEI interface for connecting peripheral data devices to TETRA mobile terminals, in effect enabling them to use the TETRA network as the connectivity solution.
Although PEI offers a number of service options~\cite{TETRA-PEI,TETRA-SCADA}, in reality manufacturers typically provide only the default option, which is the SDS.
This service has limited capabilities and provides no performance guarantees
; its main advantage is that it is accessed using simple AT commands.
Project 25 (P25) is a suite of standards developed in North America targeting public safety and emergency services~\cite{P25}, featuring a lot of analogies to TETRA, both in terms of the operational principles and of the offered services; in particular, the support for data services is more limited to the one of TETRA.
An interesting feature of P25 is that supports also the analogue mode of operation.
We also note some other relevant standards for public safety communications -- P25 and  Digital mobile radio (DMR)~\cite{DMR}, which are fairly similar to TETRA. 
For the sake of completeness, we mention other standards for public safety communications that are similar to TETRA which are TETRAPOL, P25 (originating in North America), Digital mobile radio (DMR)~\cite{DMR}., and Next-Generation Digital Narrowband (NXDN).

Another competitor in the domain of public safety communications is LTE, which introduced relevant features and services (e.g., Push-To-Talk, sidelink mode that is analogous to DMO in TETRA) starting from Release 13~\cite{Rel13}.
A major advantage of LTE 
is the native support for broadband data services and the commercial availability of LTE modems and development boards that enable straightforward implementation of customized IoT applications.
However, at the moment of writing, the commercially available LTE devices supporting sidelink mode are foreseen for automotive use cases, featuring large form factors and high energy consumption, which makes them unsuitable for the use-cases involving first responders.

A recent trend is to enable interconnection of TETRA and LTE networks.
The underlying idea is to connect the two technologies at the network core, where currently several options are under consideration: VPN tunnel over LTE core connecting TETRA software clients to the core, connecting two network cores over gateways, or a fully integrated solution featuring a single core, unified switching and user management~\cite{TCCA}.

To our knowledge, there are no prior works addressing design of a remote health monitoring system over TETRA links.
The closest work to ours can be found in~\cite{TETRA-SCADA,TETRA-SDS-perf,TETRA-GPS}.
The work in~\cite{TETRA-SCADA} considers use of a TETRA SDS service to connect remote SCADA units to the supervision center and centers on conversion of SCADA packets and character format to make it compatible for the transmission over PEI. 
The work~\cite{TETRA-SDS-perf} presented measurements of delay and outage performance of SDS transmissions among two MSs for message sizes from 10 to 190~bytes and rates of 1, 1.5 and 2~msg/s.
It was shown that SDS provides rather favorable performance when the two MSs are the only ones in the cell, while if additional 4 MSs are added to compete for the shared signalling channel resources, the performance rapidly degrades.
The evaluation performed in this work assumes a larger number of MSs running the IoT application, wider range of message generation rates and significantly larger population of devices competing for the shared resources.
The work~\cite{TETRA-GPS} performed measurement of round-trip times for SDS messages generated by two MSs every 5~s, when both were stationary, as well as when one of them was moving with a high-speed on the highway; again, our study involves a much higher number of MSs and varied message generation rates.
We also note the work~\cite{TETRA-SDS-perf-DMO}, which evaluated the performance of SDS in a railway use-case in a DMO mode, which is not in the focus of this paper.

Finally, we mention the Next Generation First Responder Apex Program, a US-government initiative that published a standardized framework for integration of the existing first-responder infrastructures with novel technological solutions~\cite{NGFR}.
The framework addresses numerous potential applications, including the remote-health monitoring.
The vision outlined in~\cite{NGFR} stresses modularity as the key principle for realization of advanced public-safety systems.
In this regard, the system studied in this paper follows these recommendations. 

\section{System Architecture and Requirements}
\label{sec:architecture}

The architecture of the considered solution is depicted in Fig.~\ref{fig:system}.\footnote{The design, development and implementation of the studied system is the topic of the currently running  Horizon 2020 project SIXTHSENSE https://sixthsenseproject.eu.}
Specifically, first responders' are equipped with a smart garment that embodies non-invasive and minimally invasive sensors recording physiological parameters of the responders, such as: temperature, hearth rate and hearth-rate variability, bio-markers for fatigue, stress, dehydration, etc.
These readings are gathered by a custom-designed microcontroller, also embedded in the garment.
Over the PEI, the microprocessor drives the TETRA Mobile Station (MS) to send the gathered recordings in the uplink to the remote agent (the type of connection between the microcontroller and PEI depends on the manufacturer/device series and can be RS-232, USB, Bluetooth, etc).
The remote agent, also connected over the PEI\footnote{An alternative, not considered in this paper, is to connect to the remote agent, located in some public network (e.g. public Internet) through a Demilitarized Zone (DMZ) of the TETRA network.}, analyzes the received data and infers the first-responders' physiological condition, based on which it feeds back the notifications and warnings.
In the considered system, the feedback consists of a predefined set of tactile stimuli which, after being received by the microcontroller, incite the electrodes in the garment to convey the electrotactile feedback to the first responders.
The choice of this particular feedback channel is driven by the fact that it does not interfere with the visual and auditory modalities of the first responders, thus minimally compromising their ability to carry out the mission.

In communication terms, the system is characterized by quasi-periodic exchanges of short data.
Specifically, an individual sensor reading is a couple bytes long, thus, the size of application-level messages that aggregate multiple readings is of the order of tens to a hundred bytes.
Similarly, the set of tactile stimuli that can be unobtrusively used in the system is limited, so a couple of bytes is enough to encode an individual stimulus.
As first responders are healthy and trained professionals, it requires a significant build-up of a physiological stress and long periods before their health conditions become jeopardized.\footnote{The main concern is that, in the course of the action, the first responders do not themselves recognize that their health status has become compromised. Indeed, overexertion and physiological stress are the leading causes of undesired and even tragic consequences~\cite{FD}.}
In effect, the periodicity of the monitoring loop can be of the order of tens of seconds, if not minutes.
Moreover, since the system performs a continuous monitoring with relatively relaxed timing constraints, the reliability of delivery of an individual application-layer message is of secondary importance, and the emphasis is on guaranteeing timeliness and reliability on the message-flow level.
In this respect, the candidate metric that can be used to assess the performance is the PAoI.
In Section~\ref{sec:evaluation}, we study the system performance both using the traditional metrics, which are message delay and outage, as well as in terms of PAoI.

\section{TETRA Short Data Service}
\label{sec:TETRA}

\begin{figure}[t!] 
		\centering
		\includegraphics[width= 0.95\linewidth]{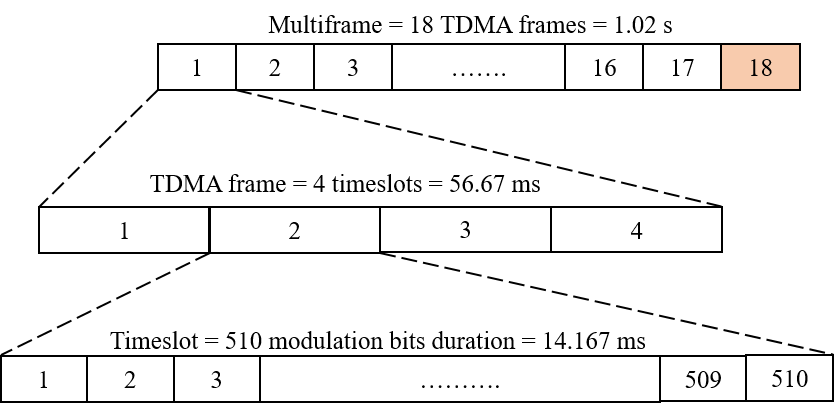}
		\caption{The TDMA structure in TETRA.  \label{tdma}}
	\end{figure}

The physical radio resourced in TETRA system are partitioned both in frequency (with carrier separation of 25 kHz) and in time. The access scheme is based on Time Division Multiple Access (TDMA), and each TDMA frame is divided in four timeslots, see Fig.~\ref{tdma}. This is further organized as 18 TDMA~frames per multiframe, where the 18th frame is used for control signaling. The time slot lasts 14.167 ms, which may be subdivided into two subslots. A physical channel in TETRA is defined by a pair of radio carrier frequencies (downlink and uplink) and a time slot number. There are two types of physical channels, Traffic Physical channel (TP) and Control Physical channel (CP). Voice and data are carried mainly by the TP, while signaling and SDS are carried by the CP. One CP channel is defined as the Main Control CHannel (MCCH), and the carrier containing the MCCH is called the main carrier. The frequency of the main carrier for the cell is broadcast by the BS, and the MCCH is located on timeslot 1 of the main carrier. SDS messages are sent over the shared MCCH.
The SDS service provides both pre-defined 16-bit messages and user-defined messages, whose length can be 16 bits (SDS type-1), 32 bits (SDS type-2), 64 bits (SDS type-3) or up to 2047 bits (SDS type-4) of application-defined data \cite{TETRA-AIR}.

    \begin{figure}[t!] 
		\centering
		\includegraphics[width=\linewidth]{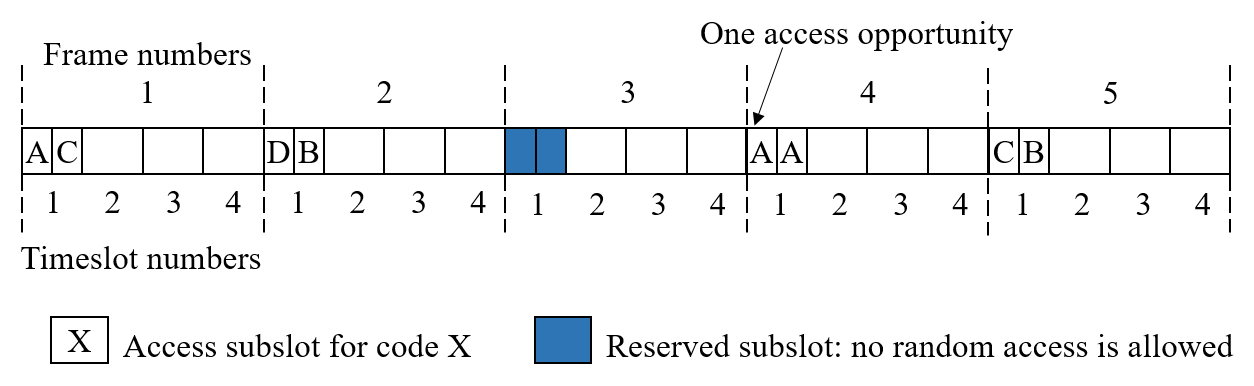}
		\caption{Structure of the uplink access subslots.  \label{codes}}
	\end{figure}

To initiate transfer of a SDS message, the MAC-layer of the TETRA MS adopts a random access procedure based on slotted ALOHA. 
Different from the conventional slotted ALOHA, radio access in TETRA system is provided only by invitation.
In each cell, MSs are grouped into four access groups, where each group is assigned an access code, 
denoted as A, B, C and D.
As shown in Fig.~\ref{codes}, the two subslots of the MCCH in each TDMA frame represent two access opportunities.
The BS marks each access opportunity with an access code and a MS can only initiate a random access request within the subslots reserved for its assigned code.
The MSs identify their designated access opportunities via the ACCESS-DEFINE message~\cite{TETRA-AIR}, a downlink control message sent by the BS at intervals decided by the operator.
The ACCESS-DEFINE message also defines the parameters for a specified access code: the Waiting Time (WT) that denotes the number of TDMA frames before an MS initiates an access retry, ranging from 1 to 15 and the Number of random access transmissions (Nu) that defines the maximum number of random access attempts by an MS, ranging from 1 to 15.

    \begin{figure}[t!] 
		\centering
		\includegraphics[width= 1\linewidth]{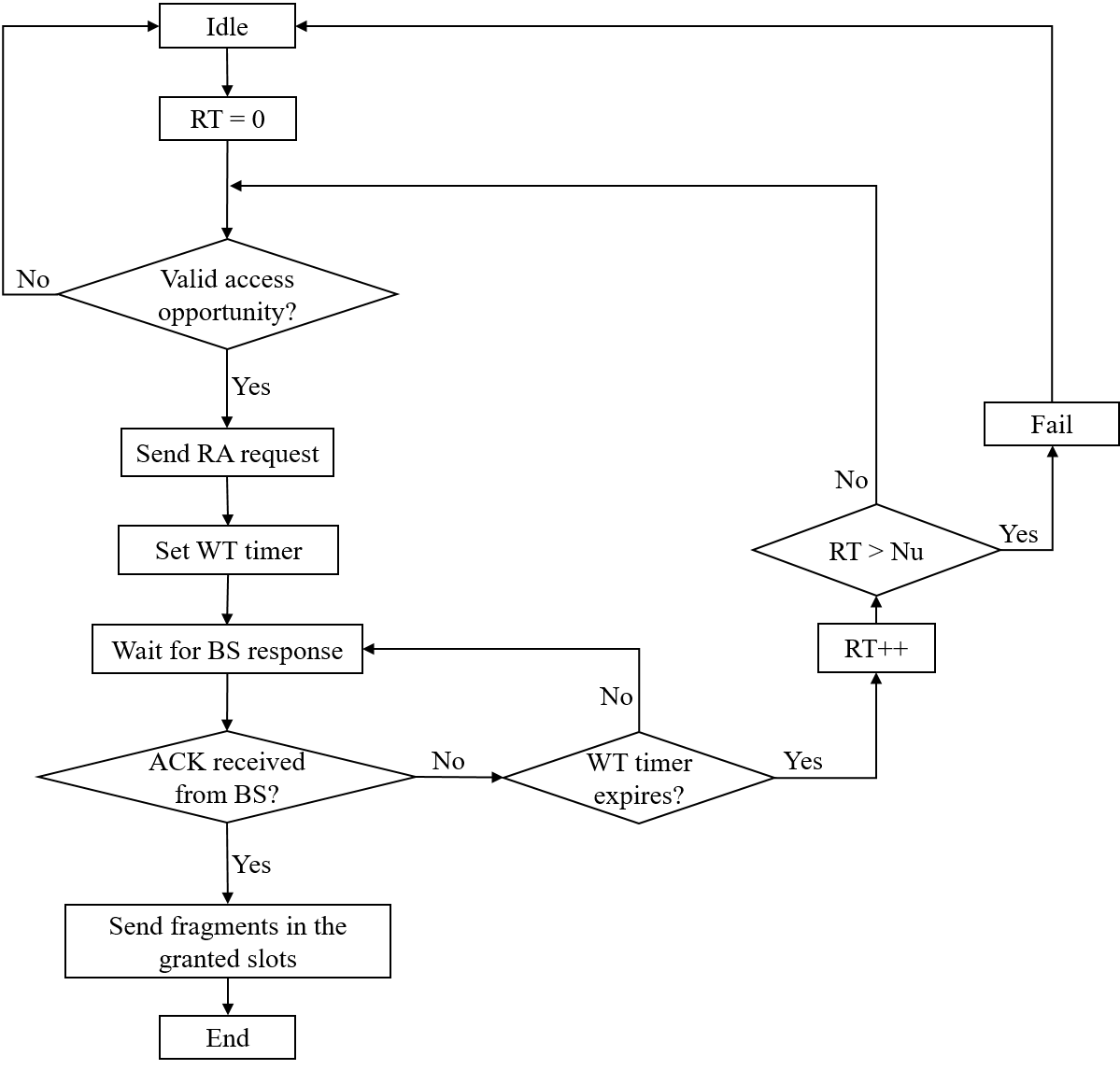}
		\caption{Flowchart for the random access procedure in TETRA.  \label{RAA}}
	\end{figure}

The sensory information gathered by the microcontroller is placed into the TETRA MS output queue as SDS fragments.
The random access procedure for transmitting the SDS fragments are described by the flowchart depicted in Fig.~\ref{RAA}, and explained as follows. Aligning with a valid access opportunity, the MS transmits the MAC-ACCESS message~\cite{TETRA-AIR} to the BS, which is random access request asking for a number of MCCH subslots to send the SDS fragments. This message also includes the first SDS fragment.
The MS then waits for the access response from the BS to approve the slot granting. If the response is not received within the WT timer, the MS shall attempt to retransmit the request in another valid access slot.
If the number of retransmission attempts exceed Nu, the transmission is considered to be failed and the MS falls back to the idle state.
Otherwise, the MS receives the MAC-RESOURCE message~\cite{TETRA-AIR} from the BS indicating a successful random access and including the corresponding reserved slots, which are marked by the BS as unavailable for random access for other MSs. The reserved access granted to the MS occupy successive subslots of the MCCH (first time slot), except that the MS should jump over frame~18. Failing to deliver the MAC-ACCESS message or receive the MAC-RESOURCE message within the time WT can be due to either collisions or poor channel quality. The transmitted SDS fragments are not numbered and are sent in sequence.
If a single transmission error occurs, the whole SDS message fails, and the MS repeats the whole procedure up to a specific retransmission limit. Upon successfully receiving the fragments and constructing the whole SDS message, the BS sends an acknowledgement (ACK) to the transmitter.
If the ACK is not received within a pre-defined interval, the MS has to retransmit the whole message again.
The BS reserves a number of downlink subslots on the MCCH to transmit the received SDS message to the destination (the remote agent in the proposed system).

\section{Evaluation}
\label{sec:evaluation}
    \begin{figure}[t!] 
		\centering
		\includegraphics[width= 1\linewidth]{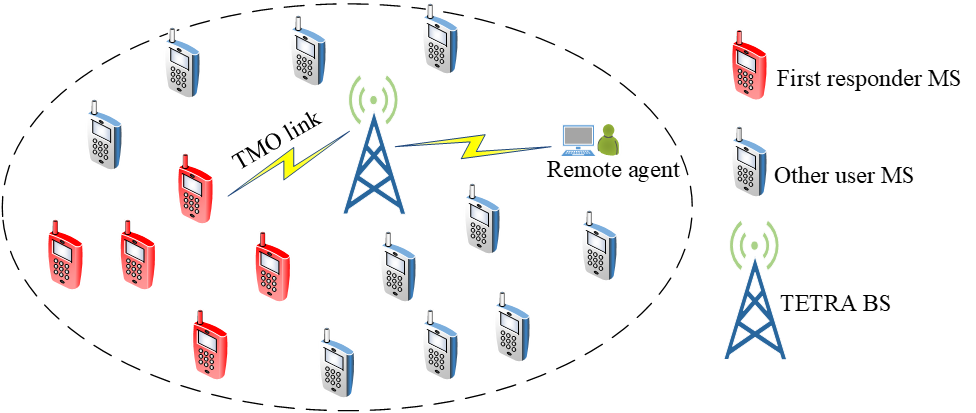}
		\caption{Graphical representation of the simulated TETRA network.  \label{netw}}
	\end{figure}


\begin{table}[t!]
		\centering
		\caption{Evaluation parameters.}
		\label{t1}
		\begin{tabular}{ll}
			\toprule
			Parameter & Value \\
				\midrule
		    Propagation model & Typical Urban (TU)\\
			Modulation & $\pi/4$-DQPSK\\
			Carrier frequency & \SI{400} {\mega\hertz}\\
			Size of SDS message & \SI{100}{\byte}\\
			SDS retry limit & \num {3}\\
			ACK waiting time & \num {4} TDMA frames\\
			$\lambda_{c}$ & \num {10} messages/hour/MS\\
			$\lambda_{\textrm{voice}}$ & \num {3} calls/hour/MS\\
			Call duration & Uniformly distributed [\SI{20}-\SI{40}{\second}]\\
			\bottomrule
		\end{tabular}	
	\end{table}

We developed a discrete-event simulator in MATLAB that evaluates the performance of the end-to-end transmission of SDS in a TETRA cell.
We assume a total of $\mathrm{N_{tot}} = \mathrm{N_C} + \mathrm{N_F} + 1$ users that are uniformly distributed within the cell coverage area and communicate with the TETRA BS through TMO radio links, see Fig.~\ref{netw}.
Specifically, $\mathrm{N_F} \in \{5,10,50\}$ users are the MSs of first responders, transmitting only SDS messages to the remote agent, following a Poisson process with parameter $\lambda_o$ that models the quasi-periodic reporting of sensory readings.
$\mathrm{N_C}$ users are standard MSs generating both voice and SDS traffic according to a Poisson process with parameters $\lambda_{\textrm{voice}}$ and $\lambda_c$, respectively.
The upper limit on $\mathrm{N_C}$ considered in the paper is 500, corresponding in practice to catastrophic situations in which a huge number of public safety forces are concentrated in a small area~\cite{TETRA-users-cell}.
Finally, the remote agent is a fixed terminal within the cell that generates a 1-byte SDS messages (notifications and warnings feedback) to the $\mathrm{N_F}$ users following a Poisson process with parameter $\lambda_F=\frac{\mathrm{N_F}}{\SI{60}{\second}}$.
The transmission of the SDS fragments follows the random access procedure presented in Section~\ref{sec:TETRA}, where each fragment is transmitted in a sequence of events shown in the flowchart in Fig.~\ref{RAA}.
The parameters listed in Table~\ref{t1} define the simulation environment; their values are selected according to the ETSI standards~\cite{TETRA, TETRA-AIR}, modeling real-world scenarios.

The developed model investigates the performance of the SDS transmissions from first responders in terms of the following parameters: 1) Average delay, which is defined as the average time elapsed between the generation of the SDS message and receiving the ACK. 
2) Failure probability, which is the ratio of dropped SDS messages over all generated ones; an SDS message is dropped if either the number of random access attempts exceeds Nu or the SDS retry limit is exceeded.
3) Average PAoI, which is the average value of the time elapsed from the moment that the preceding successfully received SDS message was generated up to the moment immediately prior to the successful reception of a new SDS message~\cite{AoI}. 
Each point in the presented results is obtained via extensive Monte-Carlo runs to attain a high confidence interval ($>95\%$), with each run lasting for 1000 TDMA multiframes. 
    \begin{figure}[t!] 
		\centering
		\includegraphics[width= 0.9\linewidth]{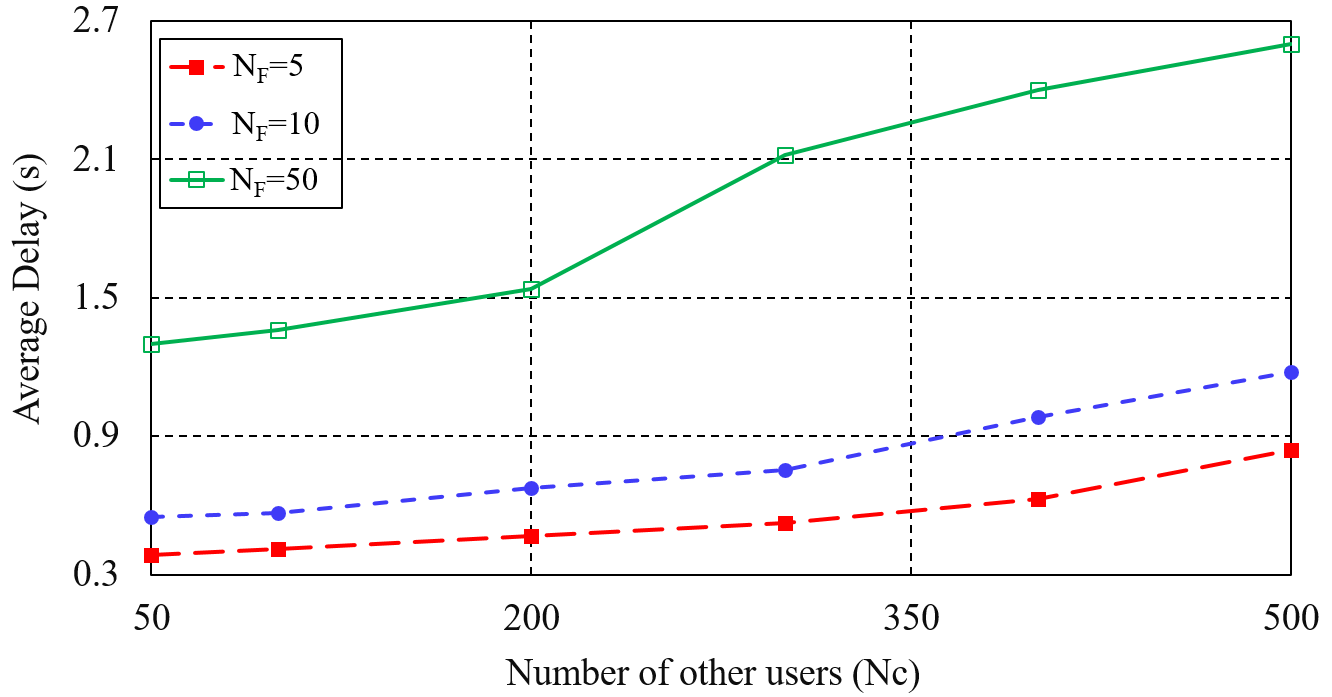}
		\caption{Average delay under different values of $\mathrm{N_F}$ with $\lambda_o=\SI{1/10}{\second}$, $\textrm{WT}=5$ and $\textrm{Nu}=5$.  \label{ad10}}
	\end{figure}
    \begin{figure}[t!] 
		\centering
		\includegraphics[width= 0.9\linewidth]{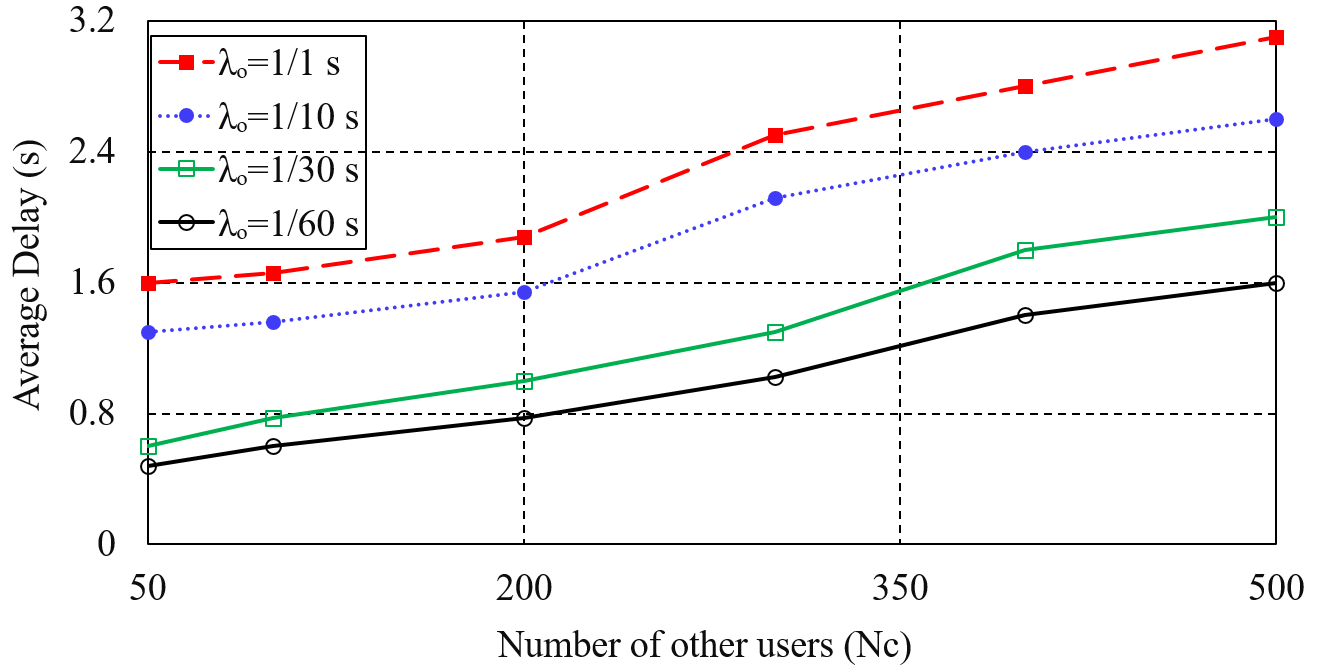}
		\caption{Average delay under different values of $\lambda_o$ with $\mathrm{N_F}=50$, $\textrm{WT}=5$ and $\textrm{Nu}=5$.  \label{ad50}}
	\end{figure}

Fig.~\ref{ad10} and Fig.~\ref{ad50} show the average delay performance as a function of the number of other users in the cell $\mathrm{N_C}$, for different values of $\mathrm{N_F}$ and $\lambda_o$, respectively.
Obviously, $\mathrm{N_C}$ has a direct effect on the average delay as all MSs share a single MCCH to transmit the signaling traffic, for both voice and SDS.
As $\mathrm{N_C}$ increases, the MCCH gets overloaded, incurring more collisions and less access opportunities.
The figures also show that the effect of $\mathrm{N_F}$ and $\lambda_o$ on the average delay is more significant than that of $\mathrm{N_C}$. This is because the first responders at the event of emergency inject higher rate of SDS messages to the network compared to that generated by the $\mathrm{N_C}$ users.
For instance, at $\lambda_o=\SI{1/10}{\second}$ and $\mathrm{N_F=5}$, the average delay increases by $34\%$ when $\mathrm{N_C}$ grows from 200 to 400, while the change is $58\%$ when $\mathrm{N_F}$ grows from 5 to 10 with $\mathrm{N_C}=200$ users. Specifically, the increase in $\mathrm{N_F}$ increases the competition to get access opportunity and an MS terminal has to wait longer before starting an SDS transmission. On the other hand, the increase in $\lambda_o$ directly impacts the queuing time at the output buffer of each MS, impacting the delay more than radio transmission impairments and access collisions.
    \begin{figure}[t!] 
		\centering
		\includegraphics[width= 0.9\linewidth]{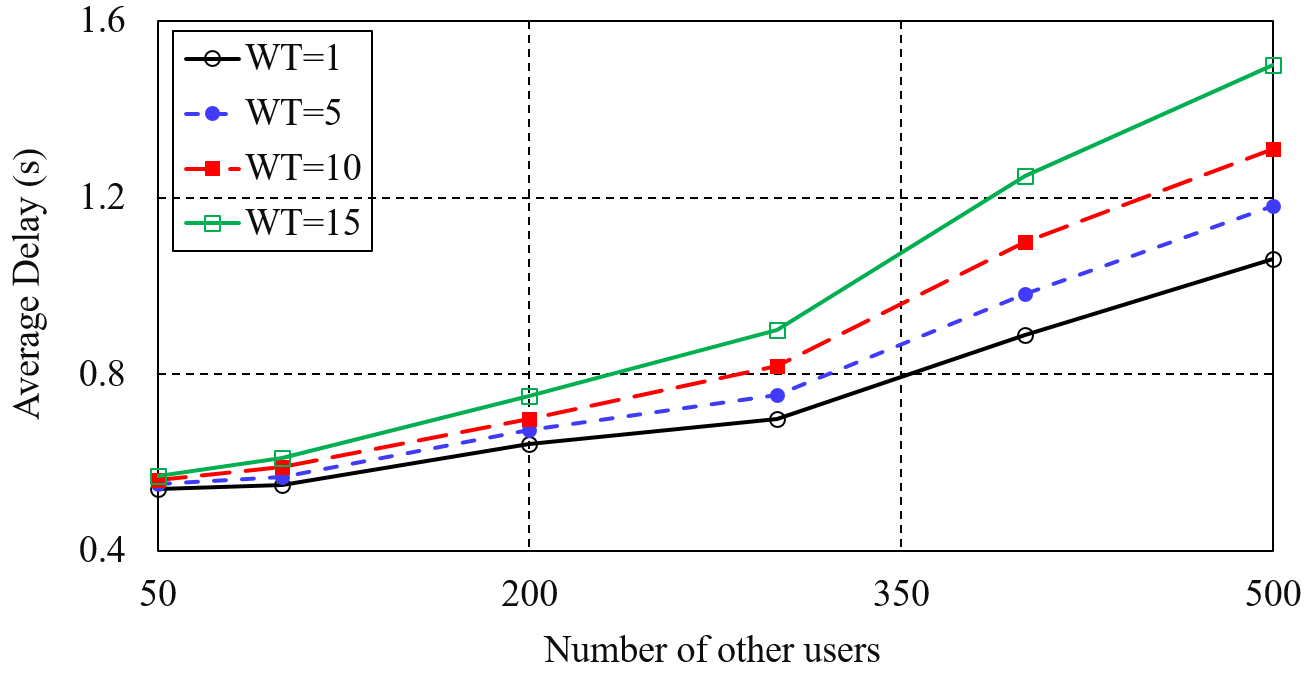}
		\caption{Average delay under different values of WT with $\lambda_o=\SI{1/10}{\second}$, $\mathrm{N_F}=10$ and $\textrm{Nu}=5$.  \label{WT}}
	\end{figure}
    \begin{figure}[t!] 
		\centering
		\includegraphics[width= 0.9\linewidth]{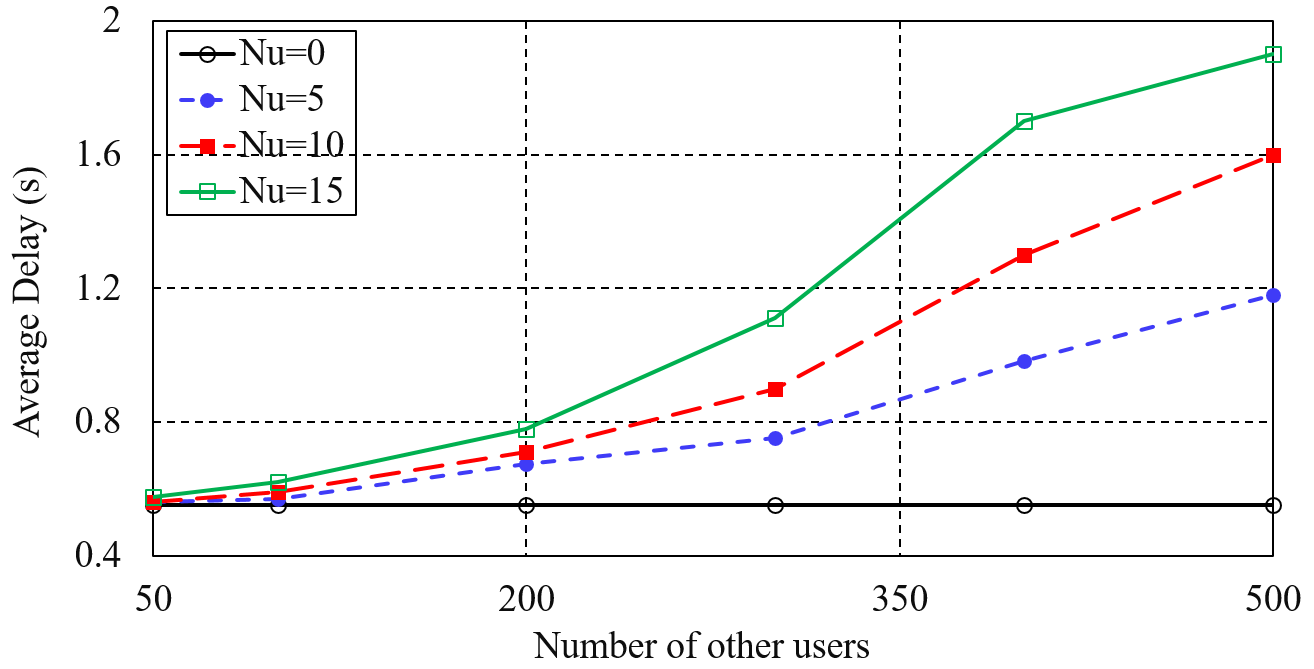}
		\caption{Average delay under different values of Nu with $\lambda_o=\SI{1/10}{\second}$, $\mathrm{N_F}=10$ and $\textrm{WT}=5$.  \label{Nu}}
	\end{figure}

In Fig.~\ref{WT} and Fig.~\ref{Nu} we investigate the effect of the waiting time WT and number of random access attempts Nu, respectively, on the average delay.
Although increasing WT would decrease the collision probability, the accessing user has to wait more time before beginning a new access retransmission, thus the average delay increases accordingly. With the considered values of $\mathrm{N_F}$ and $\lambda_o$ in Fig.~\ref{WT}, the average delay increases by $28\%$ when the value of WT changes from 5 to 15 with $\mathrm{N_C}=400$ users. The value of Nu has a more pronounced effect on the average delay performance compared to WT, 
as demonstrated by Fig.~\ref{Nu}. While the average delay increases by $28\%$ when WT changes from 5 to 10, it is increased by $72\%$ when Nu increases from 5 to 10 at the same cell population. Note that the results for $\mathrm{Nu}=0$ in essence show only the effect of a single instance of the waiting time for the messages that were successfully delivered.  
    \begin{figure}[t!] 
		\centering
		\includegraphics[width= 0.9\linewidth]{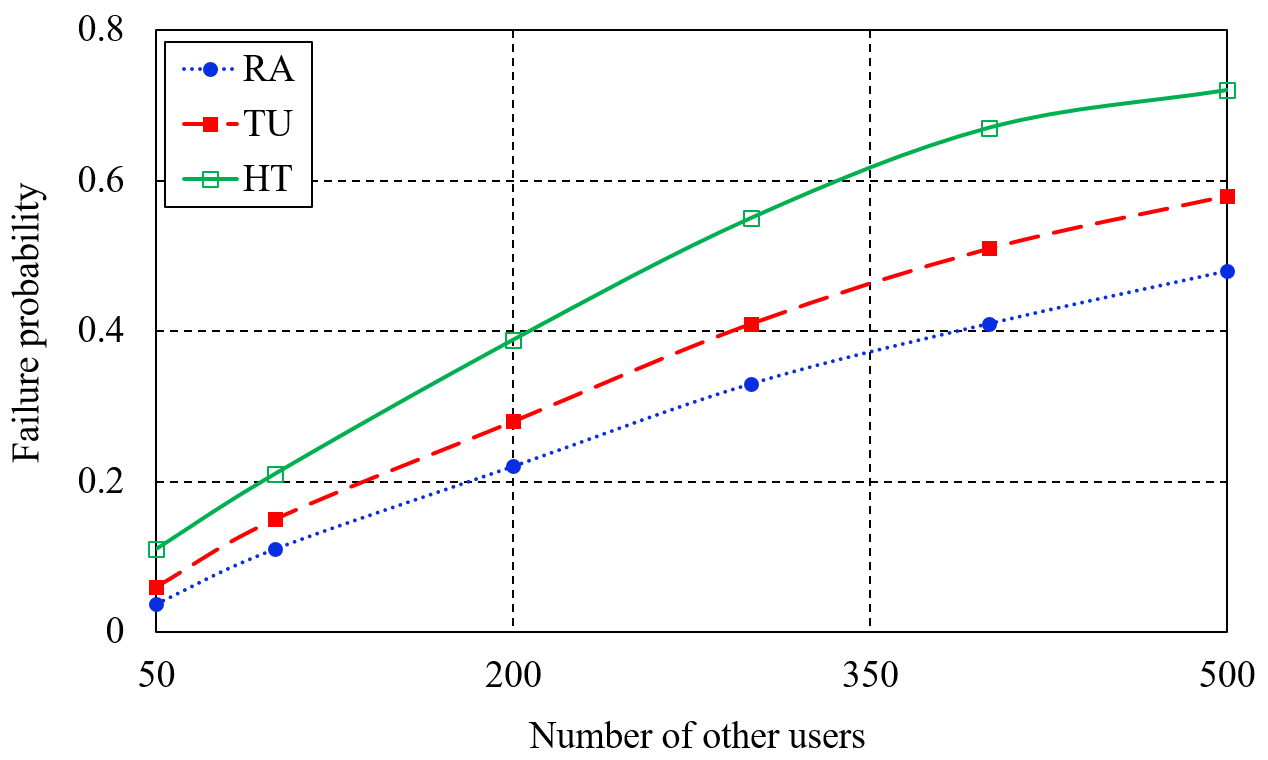}
		\caption{Failure probability under different propagation models with $\mathrm{N_F}=10$, $\lambda_o=\SI{1/10}{\second}$, $\textrm{WT}=5$ and $\textrm{Nu}=5$.  \label{fb}}
	\end{figure}

Fig.~\ref{fb} investigates the effect of the propagation environment on the failure probability.
We consider three models, Rural Area (RA), Typical Urban (TU) and Hilly Terrain (HT), which are defined in \cite{TETRA-AIR} as stochastic and statistically independent tap-gain processes.
Obviously, the performance in terms of the failure probability is better in RA, intermediate in TU and worst in HT. RA represents mostly a direct path under Rice fading whose power equals to the power of the fading process affecting the main signal. HT is the most severe environment as it has highest path delay among other models, which in turn introduces a higher inter-symbol interference. With $\mathrm{N_C}=400$ users, the failure probability in HT environment is increased by $63\%$ ($31\%$) compared to RA (TU) environment. We also note that, although the failure probability of an individual SMS message tends to high values as $\mathrm{N_C}$ increases for all propagation models, the PAoI performance can be well controlled, as demonstrated later. The presented results can guide the selection of the values of the design parameters to achieve particular  requirements. For instance, in TU environment, a TETRA site with $\mathrm{N_C}~=~100$ users can support an average delay of $\SI{560}{\milli\second}$ and failure probability lower than $0.15$ for the end-to-end transmission of the monitored parameters of 10 first responders at $\lambda_o=\SI{1/10}{\second}$, when Nu and WT are lower than 5. 

\begin{figure}[t!] 
		\centering
		\includegraphics[width= 0.9\linewidth]{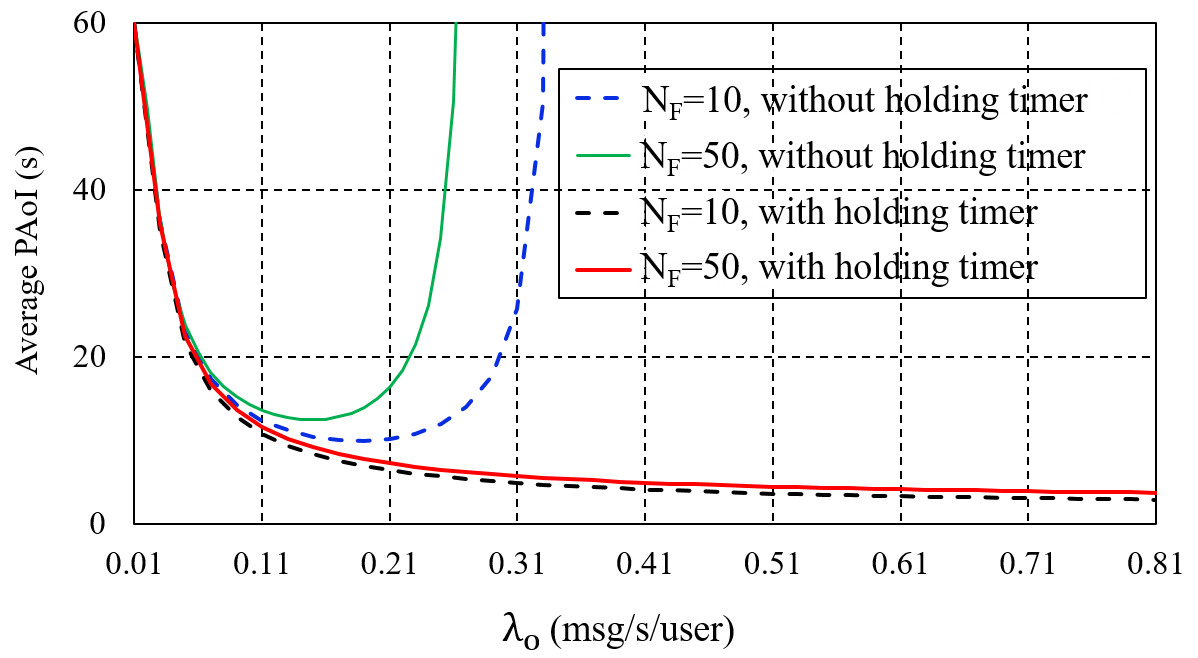}
		\caption{Average PAoI under varying  $\lambda_o$ with $\mathrm{N_C=500}$, $\textrm{WT}=5$ and $\textrm{Nu}=5$. \label{AoI1}}
	\end{figure}

So far, we considered that a packet is dropped when either WT or Nu is exceeded, otherwise, the packet remains in the output queue until the MS manages to capture access opportunity to transmit it.
However, the standard also introduces a holding timer for each packet once it enters the output queue \cite{TETRA-AIR}; when the timer elapses, the packet is dropped. The selection of the timer period directly influences the PAoI.
If the timer is too long, the packets held in the queue will prevent the newly generated ones to be transmitted, thus wasting the link time on transmission of stale information. In contrast, a short holding timer would cause packets to be prematurely dropped, again increasing the PAoI. In Fig.~\ref{AoI1}, we investigate the effect of the holding timer on the average PAoI performance under varying $\lambda_o$ with $\mathrm{N_F}$ of 10 and 50. For each point, we set the timer period to $1/\lambda_o$. As expected, low values of SDS generation rate $\lambda_o$ dictate high PAoI. As $\lambda_o$ increases, the optimized use of holding timer has a favourable effect on the average PAoI, managing to discard stale packets and enabling the remote agent to receive up-to-date ones. In this case the average PAoI decreases as $\lambda_o$ increases, allowing for timely monitoring of the health status of first responders, even when $\mathrm{N_F}$ is high. When the holding timer is not used, the PAoI becomes very large as $\lambda_o$ increases, due to the queuing and the increase in channel load and in chances of collisions caused by the stale packets. In this case, the supported range of $\lambda_o$ is roughly 0.05 - 0.25~msg/s, i.e., the reporting periods of 4 - 20~s, for which the peak age is well below one minute. 

\section{Conclusion}
\label{sec:conclusions}

In this paper, we presented a proof-of-concept study of using TETRA as the connectivity enabler for remote-health monitoring of first responders. As the system involves human-in-the-loop whose physiological status can not change abruptly, the TETRA system with its modest data capabilities, expressed in SDS service, can serve the intended purpose. In particular, the peak AoI, a measure of the freshness of the application messages exchanged in the system, follows the desired monitoring periodicity even when the network faces a huge user load.

	\bibliographystyle{IEEEtran}
\bibliography{deliverable}
	
\end{document}